\documentclass[acmsmall]{acmart}
\usepackage{bm}
\usepackage{verbatim}
\usepackage{booktabs}
\usepackage{multirow}
\usepackage{caption}
\usepackage{threeparttable}
\usepackage{graphicx}
\usepackage{float}
\usepackage{subfigure}
\usepackage[normalem]{ulem}
 \usepackage[justification=centering]{caption}
\AtBeginDocument{%
  \providecommand\BibTeX{{%
    \normalfont B\kern-0.5em{\scshape i\kern-0.25em b}\kern-0.8em\TeX}}}

\begin{document}

\title{SHA-SCP: A UI Element Spatial Hierarchy Aware Smartphone User Click Behavior Prediction Method}


\author{Ling Chen}
\orcid{0000-0003-1934-5992}
\authornote{Corresponding author.}
\email{lingchen@cs.zju.edu.cn}
\affiliation{%
  \department{College of Computer Science and Technology, Alibaba-Zhejiang University Joint Research Institute of Frontier Technologies}
  \institution{Zhejiang University}
  \streetaddress{38 Zheda Rd}
  \city{Hangzhou}
  \postcode{310027}
  \country{China}}

\author{Yiyi Peng}
\orcid{0000-0002-2572-705X}
\email{pengyiyi_zju@cs.zju.edu.cn}
\affiliation{%
  \department{College of Software Technology}
  \institution{Zhejiang University}
  \streetaddress{38 Zheda Rd}
  \city{Hangzhou}
  \postcode{310027}
  \country{China}}

\author{Kai Qian}
\orcid{0009-0004-6239-5666}
\email{22151268@cs.zju.edu.cn}
\affiliation{%
  \department{College of Software Technology}
  \institution{Zhejiang University}
  \streetaddress{38 Zheda Rd}
  \city{Hangzhou}
  \postcode{310027}
  \country{China}}
  
\author{Hongyu Shi}
\orcid{0000-0002-8625-7135}
\email{shihongyu@cs.zju.edu.cn}
\affiliation{%
  \department{College of Computer Science and Technology}
  \institution{Zhejiang University}
  \streetaddress{38 Zheda Rd}
  \city{Hangzhou}
  \postcode{310027}
  \country{China}}
  
\author{Xiaofan Zhang}
\orcid{0000-0001-5906-6563}
\email{zhangxiaofan@oppo.com}
\affiliation{%
  \department{Neutron Research Center}
  \institution{OPPO}
  \streetaddress{Huayang Street Xinchuan Technology Park OPPO Building}
  \city{Chengdu}
  \postcode{610000}
  \country{China}}

\renewcommand{\shortauthors}{Chen et al.}

\begin{abstract}
Predicting user click behavior and making relevant recommendations based on the user’s historical click behavior are critical to simplifying operations and improving user experience. Modeling UI elements is essential to user click behavior prediction, while the complexity and variety of the UI make it difficult to adequately capture the information of different scales. In addition, the lack of relevant datasets also presents difficulties for such studies. In response to these challenges, we construct a fine-grained smartphone usage behavior dataset containing 3,664,325 clicks of 100 users and propose a UI element \underline{s}patial \underline{h}ierarchy \underline{a}ware \underline{s}martphone user \underline{c}lick behavior \underline{p}rediction method (SHA-SCP). SHA-SCP builds element groups by clustering the elements according to their spatial positions and uses attention mechanisms to perceive the UI at the element level and the element group level to fully capture the information of different scales. Experiments are conducted on the fine-grained smartphone usage behavior dataset, and the results show that our method outperforms the best baseline by an average of 10.52$\%$, 11.34$\%$, and 10.42$\%$ in Top-1 Accuracy, Top-3 Accuracy, and Top-5 Accuracy, respectively.
\end{abstract}



\maketitle

\section{Introduction}
Smartphone usage behavior prediction aims to predict the user’s next smartphone usage behavior based on usage history and contexts, which is an important research area of mobile and ubiquitous computing. Based on the prediction results, the system can simplify the user’s operation process, and improve the user experience. For example, it can be used to personalize mobile apps and services \cite{liu2011personalized}, predict user needs, and provide relevant recommendations \cite{bohmer2013appfunnel,costa2012app,zhang2013nihao}. In addition, by predicting the next click, the system can speed up its response time by preloading relevant data and scheduling the CPU in advance.

Existing smartphone usage behavior prediction works are mainly divided into two categories: coarse-grained usage behavior prediction and fine-grained usage behavior prediction. Coarse-grained usage behavior prediction predicts the next app to be opened based on the user’s app opening history. Some of these early works \cite{natarajan2013app,zou2013prophet,baeza2015predicting} use probabilistic models, e.g., Markov models and Bayesian models, which ignore the long-term temporal dependences of app opening behavior sequences. To model these dependences, some works \cite{mehrotra2017understanding,do2011smartphone,bohmer2011falling,srinivasan2014mobileminer,wang2021app2vec,xia2020deepapp} use deep neural networks, e.g., long short-term memory networks (LSTMs) and recurrent neural networks (RNNs) to mine app opening behavior sequences and related contextual information, modeling them into a unified feature representation for the prediction. However, given that users tend to have multiple clicks in a single app, app opening behavior sequences are coarse-grained information about the user’s interaction with the smartphone. In these works, the multiple clicks in a single app are ignored, and the potential information in the fine-grained usage behavior of the user fails to be captured.

To this end, some fine-grained usage behavior prediction methods have been proposed \cite{zhou2021large,lee2018click}, which usually construct the sequences of clicked elements and predict the next user click behavior by deep learning models. These works either model a limited number of identified elements or model all elements in the UI through attention mechanisms at the element level. However, since the number of UI elements is huge and the positions of some elements change dynamically, it is difficult to identify and cover all elements precisely. In addition, due to the complexity and variety of the UI, the granularity of the element level perception is too fine, and it is difficult to capture large-scale information.

To address the aforementioned problems, we propose a UI element \underline{s}patial \underline{h}ierarchy \underline{a}ware \underline{s}martphone user \underline{c}lick behavior \underline{p}rediction method (SHA-SCP). SHA-SCP builds element groups and perceives the UI at the element level and the element group level to fully capture the information of different scales. The main contributions of this paper are summarized as follows:
\begin{itemize}
\item Collect and analyze a fine-grained smartphone usage behavior dataset containing 3,664,325 clicks over 1,087 apps of 100 users, which contributes new knowledge to understanding user click behaviors.
\item  Propose SHA-SCP that clusters the elements according to their spatial positions to build element groups, perceives the UI at the element level and the element group level by attention mechanisms, and fuses the information at two levels, which can capture both small-scale and large-scale information in the UI.
\item  Conduct extensive experiments on the fine-grained smartphone usage behavior dataset. The experiment results demonstrate that SHA-SCP outperforms the best baseline by an average of 10.52$\%$, 11.34$\%$, and 10.42$\%$ in Top-1 Accuracy, Top-3 Accuracy, and Top-5 Accuracy, respectively.
\end{itemize}

\section{Related Work}

In this section, an overview of works related to smartphone usage behavior prediction is provided, including coarse-grained usage behavior prediction and fine-grained usage behavior prediction.

\subsection{Coarse-Grained Usage Behavior Prediction}

Previous work has extensively investigated coarse-grained usage behavior prediction methods that are usually based on the user’s app opening history to predict the next app to be opened. Some of these early works are based on probabilistic models (e.g., Markov models \cite{gambs2012next,chen2014nlpmm,huang2012predicting,liao2013feature} and Bayesian models \cite{zou2013prophet,baeza2015predicting,shin2012understanding,natarajan2013app}). For example, Natarajan et al. \cite{natarajan2013app} used a Markov model to model app opening behavior sequences, and predicted the next app with the first-order transition probability. Zou et al. \cite{zou2013prophet} proposed a lightweight Bayesian model, which improves the prediction accuracy while saving computing resources. Baeza-Yates et al. \cite{baeza2015predicting} used app opening behavior sequences in the recent time window to predict the next app using a tree-enhanced naive Bayesian network. However, these methods cannot capture the long-term temporal dependences of app opening behavior sequences.

With the development of deep learning, methods based on deep learning models, e.g., LSTM \cite{xu2020predicting,lee2019app} and RNN \cite{xia2020deepapp}, begin to emerge. These models are designed to capture temporal dependencies in sequence data, which makes them particularly suitable for modeling app opening behavior sequences. These methods take not only app opening behavior sequences but also context information into consideration \cite{karatzoglou2012climbing,liu2013survey,liu2016predicting,zheng2010collaborative,zhu2012mining}. For example, Xu et al. \cite{xu2020predicting} proposed a general prediction model based on LSTM, which converts time series dependencies and contextual information into a unified feature representation to predict the next app. Lee et al. \cite{lee2019app} applied a stacked LSTM architecture to capture the temporal dependences of historical app opening behavior sequences. Jiang et al. \cite{jiang2019using} considered the temporal context in the prediction based on the most similarly app opening cases in history. Zhao et al. \cite{zhao2019appusage2vec} modeled apps by considering the contribution of different apps based on the temporal context. Wang et al. \cite{wang2021app2vec} introduced temporal and spatial contexts and used the Dirichlet process to capture when and where the app opening behavior occurs and what type of semantics it belongs to. Xia et al. \cite{xia2020deepapp} developed an RNN based model through multi-task learning, which introduces a location prediction task to learn the spatial-temporal relationships, and predicts both the next app and the location where the app opening behavior occurs. Shen et al. \cite{TMC2023} developed a deep reinforcement learning based prediction model and introduced a context representation method for the complex contextual environment, including temporal and spatial contexts.

To better capture the contextual information related to the user’s historical app opening behaviors, graph embedding methods were applied to the next app prediction \cite{chen2019cap,ouyang2022learning,yu2020semantic,zhou2020graph}. For example, Chen et al. \cite{chen2019cap} proposed CAP, a context-aware personalized prediction algorithm, which learns node embeddings from three subgraphs, App-Location, App-Time, and App-App type, and introduces user personalized preferences for the next app prediction. Yu et al. \cite{yu2020semantic} used graph neural networks to learn node embeddings on App-Location-Time graphs by considering app, time, and location as nodes and their co-occurrence relationships as edges. Zhou et al. \cite{zhou2020graph} proposed AHNEAP, which converts the user’s app opening history into heterogeneous graphs with attributes and achieves end-to-end prediction by learning app, location, and time embeddings. Ouyang et al. \cite{ouyang2022learning} proposed DUGN, a dynamic graph neural network, to learn effective app and user embeddings in App-User dynamic graphs by using a graph attention mechanism.

However, since users tend to have multiple clicks in a single app, the sequences of app opening behaviors are coarse-grained information about the user’s interaction with the smartphone, and such coarse-grained usage behavior prediction methods ignore the multiple clicks in a single app and cannot capture the potential information in the user’s fine-grained usage behavior.

\subsection{Fine-Grained Usage Behavior Prediction}
The user click behavior, as the smallest unit of the user’s interaction with the smartphone, contains a wealth of information. Fine-grained usage behavior prediction usually exploits the user’s historical click behavior to predict the next user click behavior based on deep learning models \cite{lee2018click,zhou2021large}. For example, Lee et al. \cite{lee2018click} constructed a sequence of clicked elements based on identified elements and proposed a next user click behavior prediction method with LSTM. However, due to the huge number of UI elements and the change of their positions, it is difficult to get precise identification of all elements. Zhou et al. \cite{zhou2021large} proposed a Transformer-based method to model user click behaviors. Transformer \cite{vaswani2017attention},  a model that uses self-attention mechanisms, has been shown to be very effective for a variety of tasks, especially those involving sequence data. In this method, the features of elements are extracted based on their text, positions, and other attributes, and the representations of elements in the UI are updated through attention mechanisms at the element level. However, UIs are often organized in a hierarchical pattern, with elements grouped according to their spatial positions. The granularity of UI perception at the element level is too fine to capture large-scale information, which may make the model’s characterization capability insufficient.

In this work, we propose a UI element spatial hierarchy aware smartphone user click behavior prediction method, which builds element groups by clustering the elements according to their spatial positions and uses attention mechanisms to perceive the UI at the element level and the element group level to fully capture the information of different scales in the UI.

\section{Dataset}

In this section, firstly the collection of a fine-grained smartphone usage behavior dataset is introduced. Secondly, the data privacy concerns and processing related to the dataset are provided. Finally, the results of the data analysis performed on the dataset are presented.

\subsection{Data Collection}

In order to collect fine-grained smartphone usage behavior data, we developed a data collection app specifically for phones with Android 10.0 or higher, as well as for Huawei phones with HarmonyOS 2.0 or higher. When a click occurs, the app records the UI where the click occurs, the clicked element, and the contextual information. For each UI, we collect its original view hierarchy and elements whose properties of \textit{clickable} or \textit{isLongClickable},  \textit{visible-to-user}, and \textit{enable} are set to true. For each element, we collect its text, type, and position attributes. For the text attribute, we first check the \textit{text} property. If it is absent or empty, we check the \textit{content-desc} property. The \textit{resource-id} property is considered as the last resort. The type attribute is extracted from the \textit{class} property, and the position attribute is extracted from the \textit{bounds} property. For the clicked element, we collect its text, type, and position attributes by the aforementioned way. We do not collect the specific position of the clicked element, but instead collect the border position. Meanwhile, the contextual information, e.g., the click time and the app where the click occurs, is also collected.

\begin{figure}[t]
  \centering
    	\begin{minipage}{0.49\linewidth}
    	    \centering
            \includegraphics[width=0.82\linewidth,]{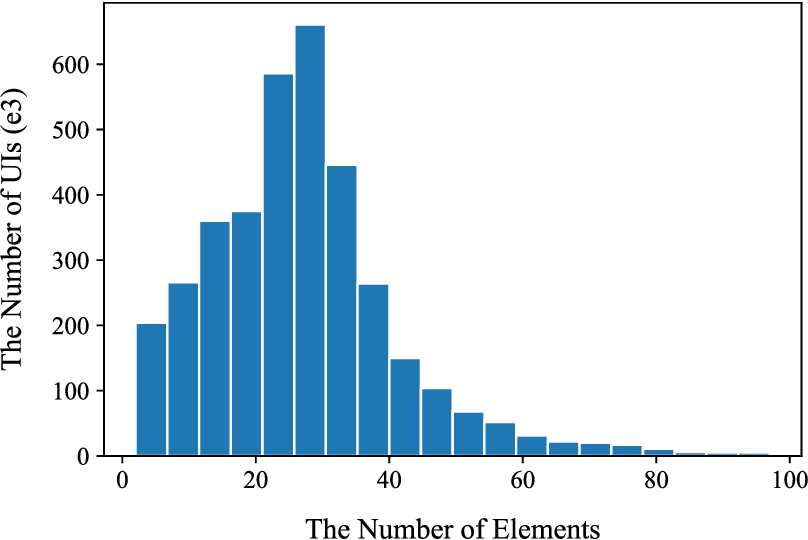}
            \caption{The distribution of the number of elements in a UI.}
    \label{fig:Ele_in_UI}  
     \end{minipage}%
		\begin{minipage}{0.49\linewidth}
			\centering
           \includegraphics[width=0.79\linewidth]{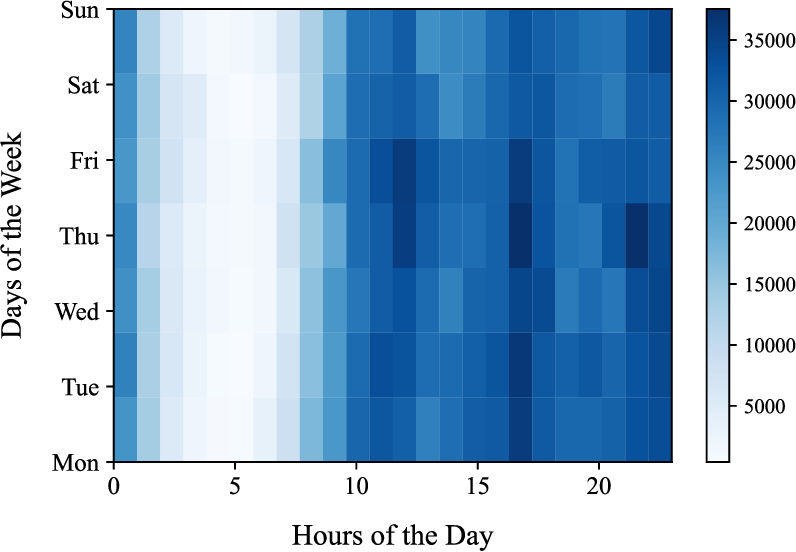}
            \caption{The distribution of clicks across the hour of the day and the day of the week.}
\label{fig:Click_Dis}  
  \end{minipage}%

  \Description{}
\end{figure}

\begin{figure}[t]
  \centering
  \includegraphics[width=0.87\linewidth]{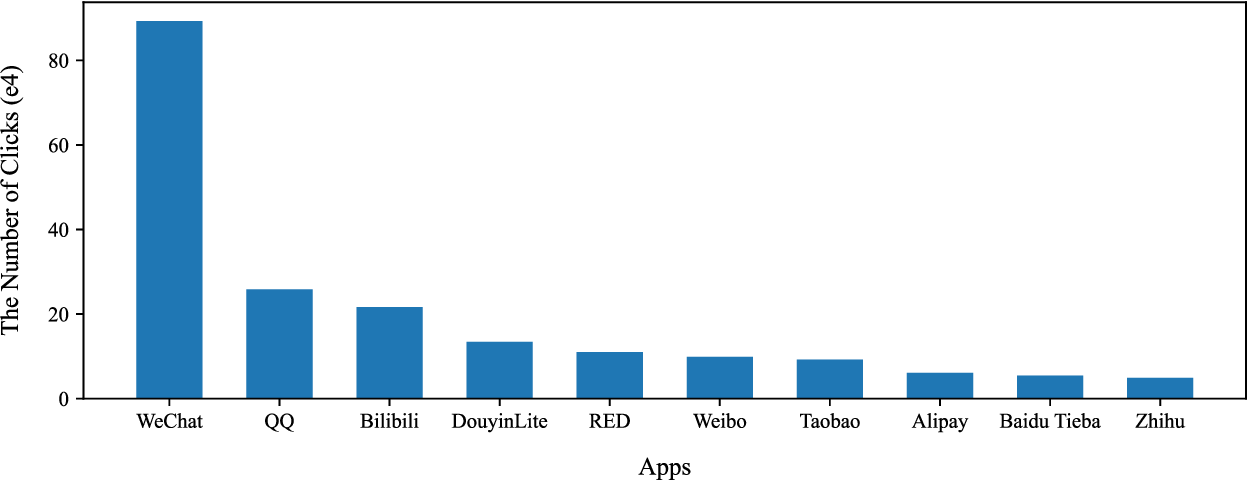}
  \caption{The distribution of clicks on the Top-10 apps.}
  \label{fig:Top_10}  
  \Description{}
\end{figure}

\begin{figure}[t]
  \centering
  \includegraphics[width=1.0\linewidth]{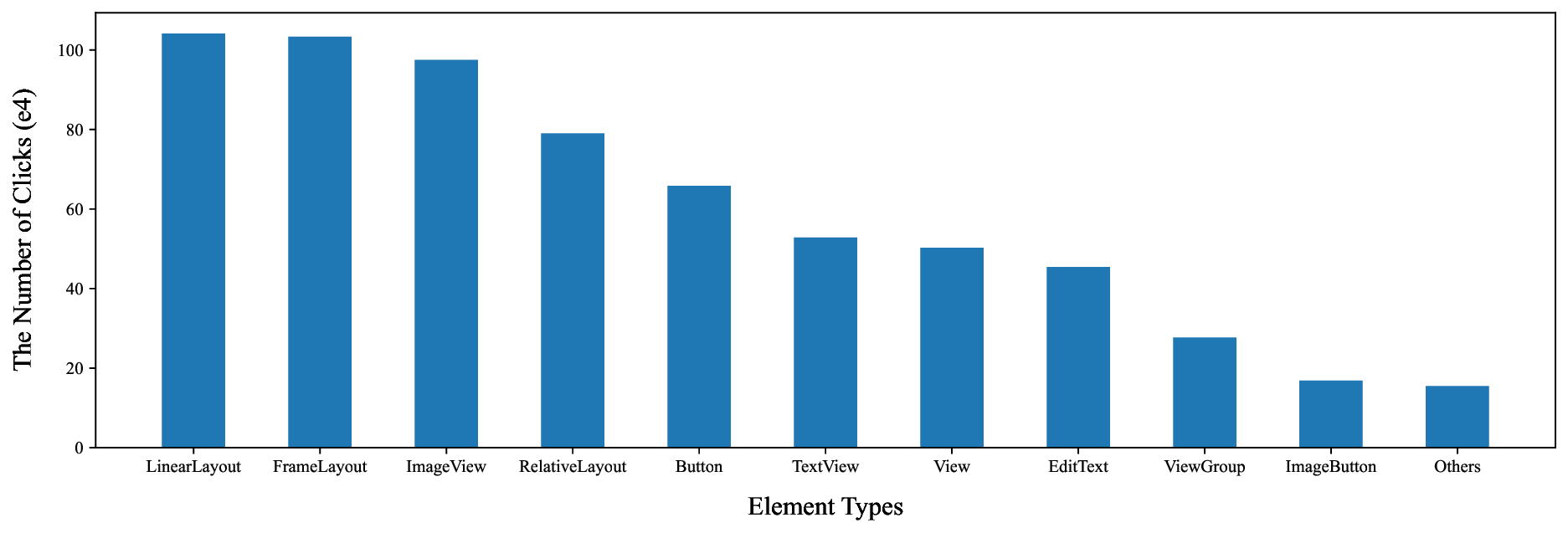}
  \caption{The distribution of the clicks of different element types.}
  \label{fig:control_count}  
  \Description{}
\end{figure}

We recruited 100 participants at a university for a fee of CNY 500/participant, and collected their fine-grained smartphone usage behavior data for 30 days. At the beginning, participants were asked to fill in questionnaires to collect basic information, including gender and age. Before the collection started, participants installed the data collection app on their daily-use smartphones and authorized it so that the app can work properly. During the collection process, participants kept the data collection app running continuously in the background of the smartphone.

Ultimately, we collected a fine-grained smartphone usage dataset containing 3,664,325 clicks over 30 days of 100 participants (47 females and 53 males, aged between 18 and 30).

\subsection{Data Privacy Concerns and Processing}
To protect privacy, we have designed measures in the whole process of data collection.

Before data collection, we clearly informed the participants of the collected data and method and explained in detail the privacy concerns. Finally, each participant signed a data authorization form, which primarily serves to clarify the utilization of data and the acquisition of usage rights for the data.

During data collection, we collected data through offline export instead of online transmission and only a unique ID was used as the identifier of each participant. When a clicked element has a text attribute and is not empty, we filter out all digits in the text and only keep the first three characters.

After data collection, we invited participants to our lab for data export and performed data encryption on site. Specifically, we use a pre-trained Sentence-BERT model [27] to embed all texts. The pre-trained Sentence-BERT model has nonlinear activation functions and feature extraction layers, which can ensure the embedding process is irreversible. Finally, to prevent restoring the text by judging the clicked keyboard position, we filter out all clicks on the mobile keyboard.

Ultimately, we presented the participants with the encrypted data. Upon their confirmation, we deleted the original data and only retained the encrypted data.

\subsection{Data Analysis}

Fig. \ref{fig:Ele_in_UI} shows the distribution of the number of elements in a UI. We can find that the number of UIs with 20 to 30 elements is the largest, accounting for 34.04$\%$. The number of elements in a UI is relatively large, with the mean value of 27 and the standard deviation of 15, which makes it a great challenge to predict the next element to be clicked from candidates.

Fig. \ref{fig:Click_Dis} shows how clicks are distributed across the hour of the day and the day of the week. The shade of color shows the number of clicks. We can find that participants tend to use smartphones in the afternoon and the evening rather than in the morning. In particular, the distribution of clicks is dense at the 17$^{th}$ hour of weekdays. Since participants are all university students, the reason may be that it is the end time of their daily course.

Fig. \ref{fig:Top_10} shows the distribution of clicks on the Top-10 apps. We can find that WeChat has the most clicks, accounting for 24.36$\%$ of all clicks. Meanwhile, clicks of the Top-10 apps account for 54.72$\%$ of all clicks, which means there are a vast number of long-tail apps.

Fig. \ref{fig:control_count} shows the distribution of the clicks of different element types. Our dataset contains a total of 71 different element types. We can find that despite having a large number of element types, the Top-10 most frequently clicked element types account for 97.65$\%$ of the total clicks. Among them, LinearLayout is the most frequently clicked element types, accounting for 15.82$\%$ of the total clicks.

\section{Model}

In this section, firstly the relevant notations and the user click behavior prediction task are defined. Secondly, the key technical novelty and general architecture of our method are introduced. Finally, each module of SHA-SCP is presented in detail.

\subsection{Problem Formulation}
We define relevant notations and formulate the user click behavior prediction task as follows.

 \textbf{Definition 1. UI.}  A UI is regarded as a set of clickable elements, denoted as  U=$\left\{e_1,e_2,\ldots{,e}_{|U|}\right\}$, where $e_1,e_2,\ldots{,e}_{|U|}$ denote clickable elements in the UI.

\textbf{Definition 2. Click.} A click consists of a tuple of attributes, including the UI, the clicked element, and the contextual information where the click occurred, denoted as $c_i=\left[U_i,e_i,\ \boldsymbol{v}_i\right]$, where $U_i$ denotes the UI, $e_i$ denotes the clicked element, and $\boldsymbol{v}_i$ denotes the contextual information, which includes the day of the week, the hour of the day, and the app where the user click occurs.

\textbf{Definition 3. Click Sequence.}  A click sequence is regarded as a series of clicks, denoted as $c_{1:N}=\left\{c_1,c_2,\ldots{,c}_N\right\}$, where $c_i$ denotes the $i^{th}$ click and $i=1,\ 2,\ \ldots\ ,\ N$.

\textbf{Problem 1. User Click Behavior Prediction Task.} The user click behavior prediction task aims to predict the next element to be clicked in the given UI, based on the contextual information and the historical click sequence, formulated as $e_{N+1}=f\left(c_{1:N},U_{N+1},\boldsymbol{v}_{N+1}\right)$, where $c_{1:N}$ denotes the historical click sequence, $U_{N+1}$ denotes the UI of the ${(N+1)}^{th}$ click, and $\boldsymbol{v}_{N+1}$ denotes the contextual information of the ${(N+1)}^{th}$ click.

\subsection{Overview}

The key technical novelty of our method is its ability to capture the rich information presented at different scales in the UI by the element spatial hierarchy awareness module. This is a crucial advancement, as previous works mainly focus on the element level, neglecting potentially valuable large-scale information.

\begin{figure}[t]
  \centering
  \includegraphics[width=0.97\linewidth]{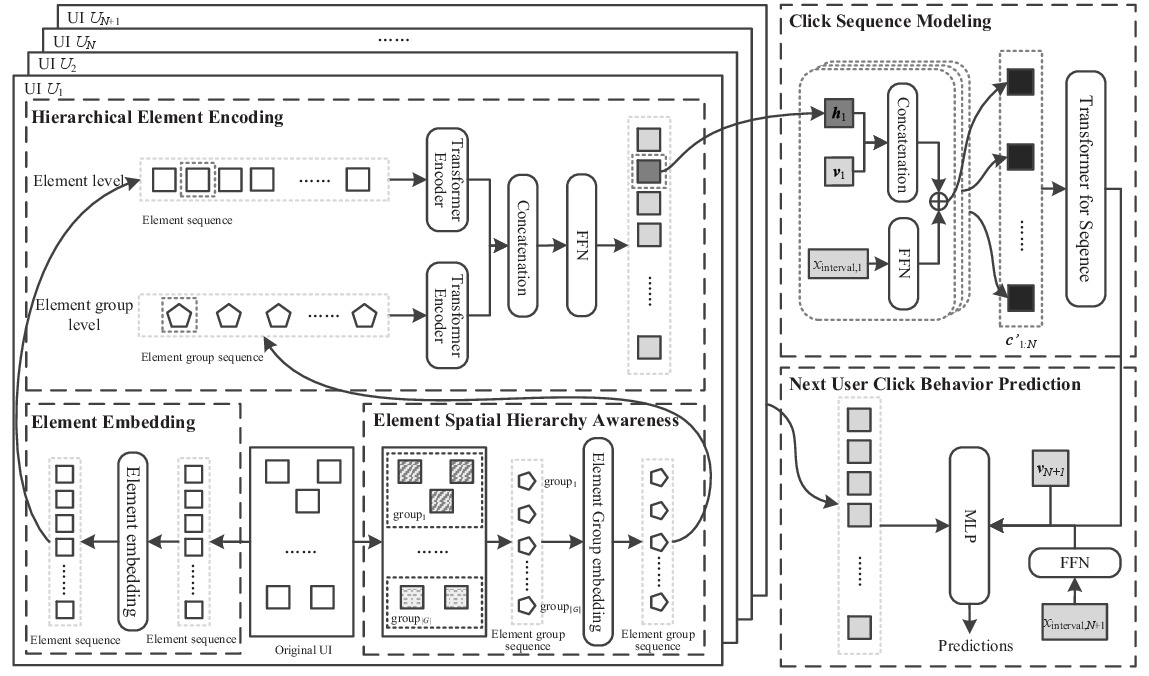}
  \caption{The overall framework of SHA-SCP.}
  \label{fig:overall}  
  \Description{}
\end{figure}

The framework of SHA-SCP is shown in Fig. \ref{fig:overall}, which consists of five modules: 
\begin{itemize}
\item Element Embedding: This module constructs a sequence of elements based on the view hierarchy of elements in the UI and obtains initial element embeddings from their own attributes.
\item Element Spatial Hierarchy Awareness: This module clusters elements according to their spatial positions to form element groups. Initial element group embeddings are obtained by aggregating the information of the elements they contain.
\item Hierarchical Element Encoding: Using attention mechanisms, this module perceives the UI at both the element level and the element group level, allowing for the capture of information at different scales.
\item Click Sequence Modeling: This module uses a Transformer to capture the impact of historical information contained in the click sequence and output a latent representation of the next clicked element.
\item Next User Click Behavior Prediction: Based on the latent representation, the UI of the ${(N+1)}^{th}$ click, and the contextual information of the ${(N+1)}^{th}$ click, this module predicts the next element to be clicked.
\end{itemize}

In Fig. \ref{fig:overall}, the original UI first proceeds to the ‘Element Embedding' module. An element sequence is constructed based on the pre-order traversal of the view hierarchy. Each element in the sequence is then represented based on its attributes (e.g., text and type attributes), outputting a represented element sequence. Meanwhile, the original UI proceeds to the ‘Element Spatial Hierarchy Awareness' module. This module first generates an element group sequence through clustering. Each element group in this sequence is then represented based on the attributes of all elements it contains, outputting a represented element group sequence.

The represented element sequence and element group sequence are then sent into a Transformer Encoder separately. The attention mechanism is employed to capture information at both the element level and the element group level. These two levels of information are then concatenated and sent into the Feed Forward Neural network (FFN), finally outputting the representations of each element.

The clicked element is then selected, and its final representation is obtained by incorporating the contextual information. The final representations of clicked elements from different UIs are composed into a click sequence, which is then sent into a Transformer. This process generates the latent representation of the next clicked element.

Lastly, the ‘Click Behavior Prediction' module predicts the next clicked element based on the latent representation, the UI of the ${(N+1)}^{th}$ click, and the contextual information of the ${(N+1)}^{th}$ click.

\subsection{Element Embedding}

For each UI, an element sequence is constructed based on the pre-order traversal of the view hierarchy. For each element, we get its initial embedding according to the attributes of text, type, and position.

For the text attribute, we use a pre-trained Sentence-BERT model \cite{reimers2019sentence} to obtain its feature $\boldsymbol{x}_{\rm text}\in\mathbb{R}^{1\times512}$. For the type attribute, we get its feature $\boldsymbol{x}_{\rm type}\in\mathbb{R}^{1\times23}$ by one-hot encoding. Then $\boldsymbol{x}_{\rm text}$ and $\boldsymbol{x}_{\rm type}$ are sent to corresponding FFNs to get the text embedding and the type embedding, whose dimensions are $d^{\rm e}$. FFNs are implemented by a linear transformation layer. For the position attribute, we first obtain the bounding coordinates of the element, then calculate the center coordinates and normalize the horizontal and vertical coordinates between 0 and 1 according to the length and width of the UI. The normalized center coordinate vector $\boldsymbol{x}_{\rm pos}\in\mathbb{R}^{1\times2}$ is sent to FFN to get the position embedding, whose dimension is ${2d}^{\rm e}$.

Following \cite{zhou2021large}, we concatenate the text embedding and the type embedding to form the content embedding of the element and sum up the content embedding and the position embedding as the initial element embedding $\boldsymbol{e}\in\mathbb{R}^{1\times{2d}^{\rm e}}$, which is formulated as:
\begin{equation}
    \boldsymbol{e}={\rm Concat}\left({\rm Embed}^{\rm text}\left(\boldsymbol{x}_{\rm text}\right),{\rm Embed}^{\rm type}\left(\boldsymbol{x}_{\rm type}\right)\right)+{\rm Embed}^{\rm pos}\left(\boldsymbol{x}_{\rm \rm pos}\right)
\end{equation}
where Concat denotes the concatenation operation and ${\rm Embed}^{\rm text}$, ${\rm Embed}^{\rm type}$, and ${\rm Embed}^{\rm pos}$ denote embedding operations on text, type, and position attributes, respectively, which are accomplished by FFNs.

\subsection{Element Spatial Hierarchy Awareness}

Due to the complexity and variety of the UI, the granularity of the element level perception is too fine to capture large-scale information. Since elements in the UI are distributed in different densities, the area with more elements may form groups of elements containing large-scale information. To this end, element groups are built by clustering the elements according to their spatial positions with the DBSCAN clustering algorithm \cite{ester1996density}. Then initial element group embeddings are obtained by aggregating the information of the elements they contain.

In the element group layer, each element group corresponds to a cluster after clustering, which may contain one or more elements. To obtain the element group text feature $\boldsymbol{x}_{\rm gtext}\in\mathbb{R}^{1\times512}$ and type feature $\boldsymbol{x}_{\rm gtype}\in\mathbb{R}^{1\times23}$, we calculate the mean values of text features and type features of all elements in the group, respectively. Then $\boldsymbol{x}_{\rm gtext}$ and $\boldsymbol{x}_{\rm gtype}$ are sent to corresponding FFNs to get the group text embedding and the group type embedding, whose dimensions are $d^{\rm g}$. To obtain the element group position feature $\boldsymbol{x}_{\rm gpos}\in\mathbb{R}^{1\times2}$, we first calculate the mean value of the normalized center coordinate vectors of all elements in the group as the coordinate vector of the group center. The group center coordinate vector is sent to FFN to get the position embedding, whose dimension is ${2d}^{\rm g}$.

The element group content embedding is obtained by concatenating the element group text embedding and type embedding. We sum up the element group content embedding and position embedding as the initial element group embedding $\boldsymbol{g}\in\mathbb{R}^{1\times2d^{\rm g}}$, which is formulated as:
\begin{equation}
   \boldsymbol{g}={\rm Concat}\left({\rm Embed}^{\rm gtext}\left(\boldsymbol{x}_{\rm gtext}\right),{\rm Embed}^{\rm gtype}\left(\boldsymbol{x}_{\rm gtype}\right)\right)+{\rm Embed}^{\rm gpos}\left(\boldsymbol{x}_{\rm gpos}\right)
\end{equation}
where ${\rm Embed}^{\rm gtext}$, ${\rm Embed}^{\rm gtype}$, and ${\rm Embed}^{\rm gpos}$ denote embedding operations on text, type, and position attributes of element groups, respectively, which are accomplished by FFNs.

\subsection{Hierarchical Element Encoding}

To update element embeddings, attention mechanisms are used to perceive the UI at the element level and the element group level to fully capture the information of different scales.

The element sequence of each UI is sent to a Transformer encoder to perceive the UI at the element level based on multi-head attention. In this process, the embeddings of elements are updated, which is formulated as:
\begin{equation}
  {\boldsymbol{e}^\prime}_{1:|U|}={\rm TransformerEncoder}^{\rm e}\left(\boldsymbol{e}_{1:|U|},\theta^{\rm e}\right)
\end{equation}
where $\boldsymbol{e}_{1:|U|}$ denotes the embeddings of the element sequence in the UI and $\theta^{\rm e}$ denotes trainable parameters.

The UI is also perceived at the element group level to fully capture the large-scale information. Element groups in a UI are sorted according to the order their elements appear in the element sequence. Similarly, the element group sequence is sent to a Transformer encoder to perceive the UI at the element group level. In this process, the embeddings of element groups are updated as well, which is formulated as:
\begin{equation}
 {\boldsymbol{g}^\prime}_{1:|G|}={\rm TransformerEncoder}^{\rm g}\left(\boldsymbol{g}_{1:|G|},\theta^{\rm g}\right)
\end{equation}
where $\boldsymbol{g}_{1:|G|}$ denotes the embeddings of the element group sequence in the UI and $\theta^{\rm g}$ denotes trainable parameters.

In order to obtain the effective representations of elements, it is necessary to take full advantage of the perception results at both the element level and the element group level. For each element, updated element and element group embeddings are fused to get its representation, which is formulated as:
\begin{equation}
\boldsymbol{h}={\rm Linear}\left({\rm Concat}\left(\boldsymbol{e}^\prime,\boldsymbol{g}^\prime\right),\beta\right)
\end{equation}
where $\boldsymbol{e}^\prime$ denotes the updated element embedding, $\boldsymbol{g}^\prime$ denotes the updated embedding of the group to which the element belongs, and $\beta$ denotes trainable parameters.

\subsection{Click Sequence Modeling}

We represent a click based on the representations of the clicked element and relevant contextual information. The representation of the clicked element obtained by the hierarchical element encoding module contains the information of the UI and the clicked element. The contextual information includes the day of the week, the hour of the day, and the app where the user click occurs.

We first get the day of the week $\boldsymbol{x}_{\rm dow}\in\mathbb{R}^{1\times7}$, the hour of the day $\boldsymbol{x}_{\rm hod}\in\mathbb{R}^{1\times24}$, and the app where the user click occurs $\boldsymbol{x}_{\rm app}\in\mathbb{R}^{1\times P}$ by one-hot encoding, where $P$ is the total number of apps. Then $\boldsymbol{x}_{\rm dow}$, $\boldsymbol{x}_{\rm hod}$, and $\boldsymbol{x}_{\rm app}$ are sent to corresponding FFNs to get their embeddings, whose dimensions are $d^{\rm v}$, $d^{\rm v}$, and ${2d}^{\rm v}$, respectively. Finally, the representation of the relevant contextual information $\boldsymbol{v}_i\in\mathbb{R}^{1\times4d^{\rm v}}$ is obtained by concatenating the embeddings of these three attributes, which is formulated as:
\begin{equation}
\boldsymbol{v}_i={\rm Concat}\left({\rm Embed}^{\rm dow}\left(\boldsymbol{x}_{\rm dow}\right),{\rm Embed}^{\rm hod}\left(\boldsymbol{x}_{\rm hod}\right),{\rm Embed}^{\rm app}\left(\boldsymbol{x}_{\rm app}\right)\right)
\end{equation}
where ${\rm Embed}^{\rm dow}$, ${\rm Embed}^{\rm hod}$, and ${\rm Embed}^{\rm app}$ denote embedding operations on the day of the week, the hour of the day, and the app where the user click occurs, respectively, which are accomplished by FFNs.

Then the representation of the click is obtained by concatenating the representations of the clicked element and relevant contextual information, which is formulated as:
\begin{equation}
\boldsymbol{c}_i={\rm Concat}\left(\boldsymbol{h}_i,\boldsymbol{v}_i\right)
\end{equation}
where $\boldsymbol{h}_i$ denotes the representation of the clicked element and $\boldsymbol{v}_i$ denotes the representation of relevant contextual information.

In order to fully consider the influence of time intervals in the click sequence, we embed the interval information between the occurred time of the click and the occurred time of the previous click, and sum up the representation of the click and the interval embedding to get the final representation of the click, which is formulated as:
\begin{equation}
{\boldsymbol{c}^\prime}_i=\boldsymbol{c}_i+{\rm Embed}^{\rm interval}\left(x_{{\rm interval},i}\right)
\end{equation}
where ${\rm Embed}^{\rm interval}$ denotes the embedding operation on the time interval, which is accomplished by FFN, $x_{{\rm interval},i}$ denotes the time interval, and $\boldsymbol{c}_i$ denotes the representation of the click.

A Transformer is introduced to capture the impact of historical information contained in the click sequence and output the latent representation of the next clicked element, which aggregates the information of the entire sequence and is formulated as:
\begin{equation}
\hat{\boldsymbol{e}}={\rm Transformer}^{\rm seq}\left({\boldsymbol{c}^\prime}_{1:N},\theta^{\rm seq}\right)
\end{equation}
where ${\boldsymbol{c}^\prime}_{1:N}$ denotes the final representations of the click sequence, $\theta^{\rm seq}$ denotes trainable parameters.

\subsection{Next User Click Behavior Prediction}

We take all elements in the UI of the ${(N+1)}^{th}$ click as candidates for the next click and calculate their clicking probabilities. The representations of elements in the UI of the $({N+1)}^{th}$ click (calculated by Equation (5)) are concatenated with the latent representation of the next clicked element (calculated by Equation (9)) and the representation of the contextual information of the ${(N+1)}^{th}$ click (calculated by Equation (6)), which are sent to a Multilayer Perceptron (MLP) and a softmax function to calculate the clicking probabilities of all elements in the UI, which is formulated as:
\begin{equation}
\alpha_k={\rm MLP}\left({\rm Concat}\left({\boldsymbol{e}^\prime}_k,{\hat{\boldsymbol{e}},\boldsymbol{v}}_{N+1}\right),\varphi\right)
\end{equation}
\begin{equation}
p_k=\frac{{\rm exp}\left(\alpha_k\right)}{\sum_{m=1}^{M}\alpha_m}
\end{equation}
where $\hat{\boldsymbol{e}}$ denotes the latent representation of the next clicked element, ${\boldsymbol{e}^\prime}_k$ denotes the representation of the $k^{th}$ element in the UI of the ${(N+1)}^{th}$ click, $\boldsymbol{v}_{N+1}$ denotes the contextual information of the ${(N+1)}^{th}$ click, $\varphi$ denotes trainable parameters, and $M$ denotes the number of elements in the UI of the ${(N+1)}^{th}$ click.

The training objective of SHA-SCP is to minimize the cross-entropy loss, which is formulated as:
\begin{equation}
L=-\frac{1}{\mu}\sum_{u=1}^{\mu}\sum_{m=1}^{M}{y_{u,m}{\rm log}\left(p_{u,m}\right)}
\end{equation}
where $\mu$ and $M$ are the number of samples in the training set and the number of elements in the UI of the ${(N+1)}^{th}$ click, respectively. $y_{u,m}$ denotes the label of the $m^{th}$ element in the UI of the ${(N+1)}^{th}$ click for sample $u$ and $y_{u,m}\in\{0,1\}$. $p_{u,m}$ is the corresponding predicted probability calculated by Equation (11).

\section{Experiments}

In this section, extensive experiments are conducted to justify the superiority of SHA-SCP. Firstly, the experimental settings are introduced. Then the comparison with baselines, the comparison with simplified models, the parameter sensitivity analysis, the feature effectiveness analysis, and the case study are presented.

\subsection{Experimental Settings}

We evaluate SHA-SCP on the fine-grained smartphone usage behavior dataset. We use a fixed-size sliding window to get click sequences, and the dataset is split into training, validation, and test sets by the ratio of 8:1:1 according to the temporal order. The dataset contains clicks over a period of 30 days from 100 users, in which a total of 3,664,325 valid clicks are collected. The minimum number of clicks per user is 9,556, and the maximum number is 100,000. The mean number of clicks per user is 36,643, with a standard deviation of 15,114. The dataset covers 1,087 apps.

SHA-SCP is implemented in Python with PyTorch \cite{paszke2019pytorch}, and the source code will be available when the paper is published. Adam \cite{kingma2014adam} is adopted as the optimizer. For the learning rate, we select it from 0.0001, 0.001, 0.01, and 0.1, and the optimal setting is 0.001. Note that, Adam optimizer can dynamically adjust the learning rate during model training. For the length of the sliding window, we select it from 0 to 10, and the optimal setting is 10. The sliding step is set to 1. For the number of the heads of multi-head attention mechanisms, we select it from 1, 2, 4, and 8, and the optimal setting is 2. The number of the layers of the MLP is set to 2. $d^{\rm e}$, $d^{\rm g}$, and $d^{\rm v}$ are set to 16, 8, and 4, respectively.

We use ${\rm Top-}k\ \rm Accuracy$, where $k$ is set to 1, 3, and 5, as evaluation metrics. ${\rm Top-}k\ \rm Accuracy$  refers to the percentage of correctly predicted samples based on the labels with ${\rm Top-}k$  predicted probabilities. Higher ${\rm Top-}k\ \rm Accuracy$  demonstrates better performance, which is formulated as:
\begin{equation}
{\rm Top-}k\ {\rm Accuracy}=\frac{\rm \#\ corrected\ prediction}{\rm \#\ test\ samples}
\end{equation}
where $\rm \#\ corrected\ prediction$  denotes the number of correctly predicted samples and $\rm \#\ test\ samples$  denotes the number of test samples.

\subsection{Comparison with Baselines}

To justify the superiority of SHA-SCP, we compare it with other user click behavior prediction methods. Among these methods, Logistic Regression (LR) \cite{lavalley2008logistic}, Support Vector Machines (SVM) \cite{hearst1998support}, and Naive Bayes (NB) \cite{rish2001empirical} are traditional machine learning methods, and LSTM \cite{hochreiter1997long}, Transformer \cite{vaswani2017attention}, and Google Model \cite{zhou2021large} are deep learning methods. For the sake of fairness, all compared methods follow the same experimental settings as SHA-SCP, and their parameters are also optimized. For Google Model, we select its learning rate from 0.0001, 0.001, and 0.01, and the optimal setting is 0.001. For the length of the sliding window, we select it from 0 to 10, and the optimal setting is 10. For the number of the heads of multi-head attention mechanisms, we select it from 1, 2, 4, and 8, and the optimal setting is 1. Other baselines are also tuned to the optimum as much as possible. The detailed descriptions of baselines are as follows:

\textbf{LR} \cite{lavalley2008logistic}, \textbf{SVM} \cite{hearst1998support}, and \textbf{NB} \cite{rish2001empirical}: LR, SVM, and NB are traditional machine learning methods. To use them on the user click behavior prediction task, we first represent elements based on their attributes of text, type, and position. Then the concatenations of the representations of elements in the UI of the ${(N+1)}^{th}$  click and the representation of the last clicked element are fed into LR, SVM, or NB to obtain the clicking probabilities of candidate elements.

\textbf{LSTM} \cite{hochreiter1997long} and \textbf{Transformer} \cite{vaswani2017attention}: LSTM and Transformer are both sequence methods. LSTM is a recurrent neural network. Transformer is a deep learning method based on self-attention mechanisms. To use them on the user click behavior prediction task, we first represent elements based on their attributes of text, type, and position and feed the click sequence into LSTM or Transformer to get the latent representation of the next clicked element. Then the representations of elements in the UI of the ${(N+1)}^{th}$  click are concatenated with the latent representation of the next clicked element. Finally, an MLP is introduced to obtain the clicking probabilities of candidate elements.

\textbf{Google Model} \cite{zhou2021large}: Google Model is a user click behavior prediction method that represents elements according to their attributes of text, type, and position and obtains the representations of elements through attention mechanisms at the element level. Given a click sequence, a Transformer is used to get the latent representation of the next clicked element, which is concatenated with the representations of elements in the UI of the ${(N+1)}^{th}$  click. Finally, an MLP is introduced to obtain the clicking probabilities of candidate elements.

Table \ref{tab:tabel2} shows the Top-1 Accuracy, Top-3 Accuracy, and Top-5 Accuracy of SHA-SCP and baselines on the fine-grained smartphone usage behavior dataset, from which we can observe the following phenomena:

1) Deep learning methods (LSTM and Transformer) outperform traditional machine learning methods (LR, SVM, and NB) on the user click behavior prediction task, which indicates the effectiveness of sequence modeling.

2) As the state-of-the-art method on the user click behavior prediction task, Google Model performs better than LSTM and Transformer, which indicates the advantages of perceiving the rich information in the UI, in addition to considering the features of the clicked element.

3) The performances of LSTM and Transformer on the user click behavior prediction task are similar. The reason is that although Transformer has a stronger ability to capture long-term dependencies, the short sequence length in this task makes it difficult for Transformer to give full play to its advantages.

4) SHA-SCP outperforms the best baseline with improvements of 10.52$\%$, 11.34$\%$, and 10.42$\%$ in Top-1 Accuracy, Top-3 Accuracy, and Top-5 Accuracy, respectively. The results justify the advantage of capturing the information of different scales by perceiving the UI at the element level and the element group level.

\subsection{Comparison with Simplified Models}

To justify the effectiveness of attention mechanisms at the element level and the element group level, we compare SHA-SCP with its simplified models. For the sake of fairness, all simplified models follow the same experimental settings as SHA-SCP, and their parameters are also optimized. The detailed descriptions of simplified models are as follows:

\textbf{SHA-SCP w/o element-attn}: SHA-SCP w/o element-attn removes the attention mechanism at the element level from SHA-SCP and only utilizes the initial element embeddings. The rest is the same as SHA-SCP.

\textbf{SHA-SCP w/o group-attn}: SHA-SCP w/o group-attn removes the attention mechanism at the element group level from SHA-SCP and only utilizes the initial element group embeddings. The rest is the same as SHA-SCP.

Table \ref{tab:tabel3} shows the Top-1 Accuracy, Top-3 Accuracy, and Top-5 Accuracy of SHA-SCP and its simplified models on the fine-grained smartphone usage behavior dataset, from which we can observe the following phenomena:

1) SHA-SCP outperforms SHA-SCP w/o element-attn, which indicates the effectiveness of using the attention mechanism to perceive the UI at the element level to capture small-scale information.

2) SHA-SCP outperforms SHA-SCP w/o group-attn, which indicates the effectiveness of using the attention mechanism to perceive the UI at the element group level to capture large-scale information.

3) The performance degradation of SHA-SCP w/o element-attn is more than that of SHA-SCP w/o group-attn, which indicates that the attention mechanism at the element level is more effective to improve the characterization capability of the model. The reason may be that the number of element groups is relatively small, which limits the effect of the attention mechanism at the element group level.

\begin{table}[t]
  \caption{Comparisons with baselines. The best results are bold and the second best results are underlined.}
  \label{tab:freq}
  \begin{tabular}{cccc}
    \toprule
   Model&Top-1 Accuracy&Top-3 Accuracy&Top-5 Accuracy\\
    \midrule
    LR	&14.11$\%$	&27.10$\%$	&35.20$\%$ \\
    SVM	&9.02$\%$	&16.60$\%$	&30.49$\%$ \\
    NB	&13.20$\%$	&18.81$\%$	&28.39$\%$ \\
    LSTM	&15.72$\%$	&28.79$\%$	&39.57$\%$ \\
    Transformer	&14.36$\%$	&28.03$\%$	&39.40$\%$ \\
    Google Model	&\underline{24.82$\%$}	&\underline{45.50$\%$}	&\underline{57.85$\%$} \\
    \textbf{SHA-SCP}	&\textbf{35.34}$\textbf{\%}$	&\textbf{56.84}$\textbf{\%}$	&\textbf{68.27}$\textbf{\%}$ \\
  \bottomrule
\end{tabular}
\label{tab:tabel2}%
\end{table}
\begin{table}[t]
  \caption{Comparisons with simplified models. The best results are bold.}
  \label{tab:freq}
  \begin{tabular}{cccc}
    \toprule
   Model&Top-1 Accuracy&Top-3 Accuracy&Top-5 Accuracy\\
    \midrule
    SHA-SCP w/o element-attn	&19.67$\%$	&36.24$\%$	&47.59$\%$ \\
    SHA-SCP w/o group-attn	&29.51$\%$	&52.84$\%$	&64.85$\%$ \\
    \textbf{SHA-SCP}	&\textbf{35.34}$\textbf{\%}$	&\textbf{56.84}$\textbf{\%}$	&\textbf{68.27}$\textbf{\%}$ \\
  \bottomrule
\end{tabular}
\label{tab:tabel3}
\end{table}

\subsection{Parameter Sensitivity Analysis}

To study the impact of several important parameters, including the length of the sliding window and the number of the heads of multi-head attention mechanisms, we conduct the parameter sensitivity analysis on the fine-grained smartphone usage behavior dataset.

\begin{figure}[h]
  \centering
		\begin{minipage}[t]{0.49\linewidth}
			\centering
			
  \includegraphics[width=0.82\linewidth,]{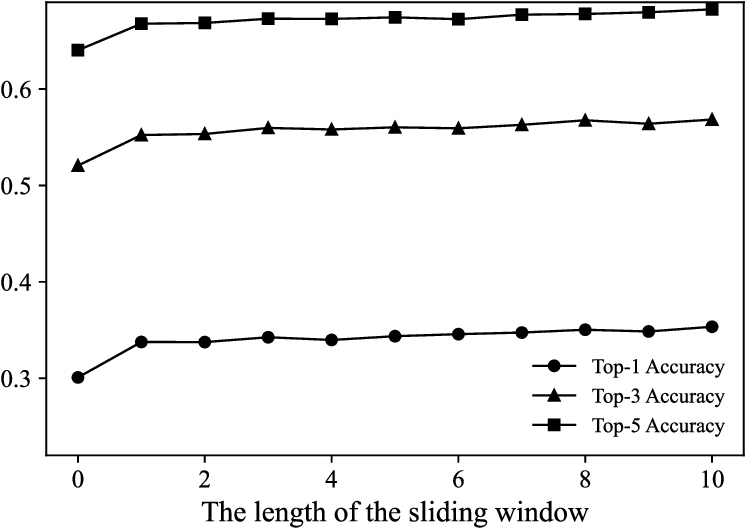}
  \caption{The impact of the length of the sliding window.}
         \label{fig:win}%
		\end{minipage}%
		\begin{minipage}[t]{0.49\linewidth}
			\centering
			
   \includegraphics[width=0.82\linewidth]{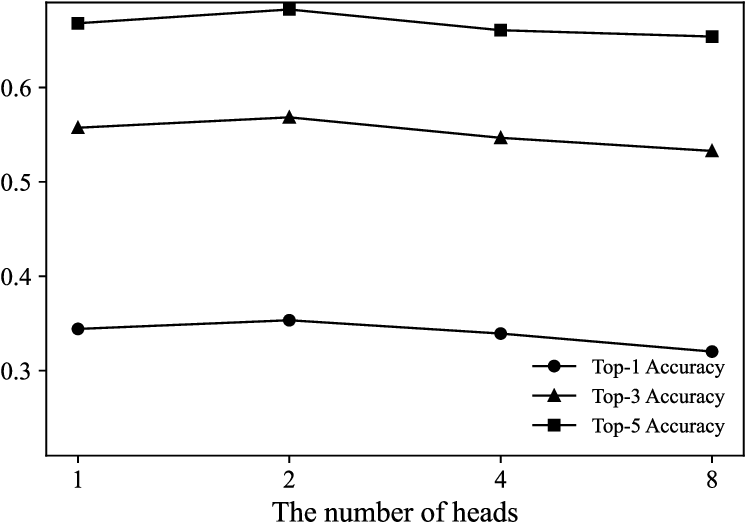}
  \caption{The impact of the number of the heads of multi-head attention mechanisms.}
         \label{fig:head}%
		\end{minipage}%

  \Description{}
\end{figure}

We evaluate the impact of the length of the sliding window, increasing it from 0 to 10 with a step size of 1. Fig. \ref{fig:win} shows the evaluation results, from which we can observe the following phenomena:

1) When the length of the sliding window increases from 0 to 1, the performance increases rapidly, which indicates the effectiveness of considering the historical click sequence.

2) When the length of the sliding window is larger than 0, with the increase of the length, the performance increases lightly, which indicates that using more historical information only has a slight effect for improving the performance.

We evaluate the impact of the number of the heads of multi-head attention mechanisms, setting it to 1, 2, 4, and 8. Fig. \ref{fig:head} shows the evaluation results, from which we can observe that with the increase of the number of heads, the performance first increases and then degrades. The best performance is achieved when the number of the heads is 2. The reason may be that when the number of heads is too small, it is difficult for the model to discover the potential information of different feature spaces, and when the number of heads is too large, the high model complexity increases the risk of over-fitting.

\subsection{Feature Effectiveness Analysis}

To justify the effectiveness of selected features, we remove selected features one by one to study the impact on the model. SHA-SCP w/o text, SHA-SCP w/o type, SHA-SCP w/o position, SHA-SCP w/o time, and SHA-SCP w/o app are models that remove the text, type, position, contextual information of time, and contextual information of app, respectively.

Table \ref{tab:tabel4} shows the Top-1 Accuracy, Top-3 Accuracy, and Top-5 Accuracy of SHA-SCP and models that remove different features on the fine-grained smartphone usage behavior dataset, from which we can observe the following phenomena:

1) SHA-SCP outperforms SHA-SCP w/o text, SHA-SCP w/o type, SHA-SCP w/o position, SHA-SCP w/o time, and SHA-SCP w/o app, which indicates the effectiveness of the selected features.

2) The performances of models with different features removed are different and the importances of features can be ranked as follows: text > position > type > contextual information of time > contextual information of app.

\begin{table}
  \caption{Comparisons with models that remove different features. The best results are bold.}
  \label{tab:freq}
  \begin{tabular}{cccc}
    \toprule
   Model&Top-1 Accuracy&Top-3 Accuracy&Top-5 Accuracy\\
    \midrule
    SHA-SCP w/o text	&29.01$\%$	&49.21$\%$	&60.78$\%$ \\
    SHA-SCP w/o type	&32.44$\%$&53.19$\%$	&64.28$\%$ \\
    SHA-SCP w/o position	&30.95$\%$	&50.77$\%$	&62.39$\%$ \\
    SHA-SCP w/o time	&33.24$\%$	&54.09$\%$	&66.27$\%$ \\
    SHA-SCP w/o app	&34.45$\%$	&55.73$\%$	&67.68$\%$ \\
    \textbf{SHA-SCP}	&\textbf{35.34}$\textbf{\%}$	&\textbf{56.84}$\textbf{\%}$	&\textbf{68.27}$\textbf{\%}$ \\
  \bottomrule
\end{tabular}
\label{tab:tabel4}%
\end{table}

\begin{table}
  \caption{The computation costs of SHA-SCP and Google Model. The best results are bold.}
  \label{tab:freq}
  \begin{tabular}{ccccc}
    \toprule
   Model&$\#$Parameters&Training time$/$epoch&Total training time&Prediction time$/$click\\
    \midrule
    \textbf{Google Model}	&\textbf{381,717}	&\textbf{2.1h}	&\textbf{6.3h} &\textbf{5.1ms} \\
    SHA-SCP 	        &595,949	&4.8h	&19.2h &6.4ms \\
  \bottomrule
\end{tabular}
\label{tab:tabel5}%
\end{table}

\subsection{Runtime Analysis}
To evaluate the computation cost, we compare the number of parameters, training time, and prediction time of SHA-SCP and Google Model on the dataset in Table \ref{tab:tabel5}. The experiments are conducted on a system equipped with a CPU Intel(R) Core(TM) i9-12900K and a GPU NVIDIA GeForce RTX 3090. SHA-SCP has more parameters than Google Model, requiring longer training and prediction time. Overall, while our method demonstrates higher prediction accuracy, the spatial hierarchy aware and hierarchical element encoding modules do increase the computation cost. Improving the efficiency of our method will be a part of future work.

\subsection{Case Study}

\begin{figure}[t]
  \centering
		\begin{minipage}[t]{0.49\linewidth}
			\centering
			
  \includegraphics[width=0.82\linewidth,]{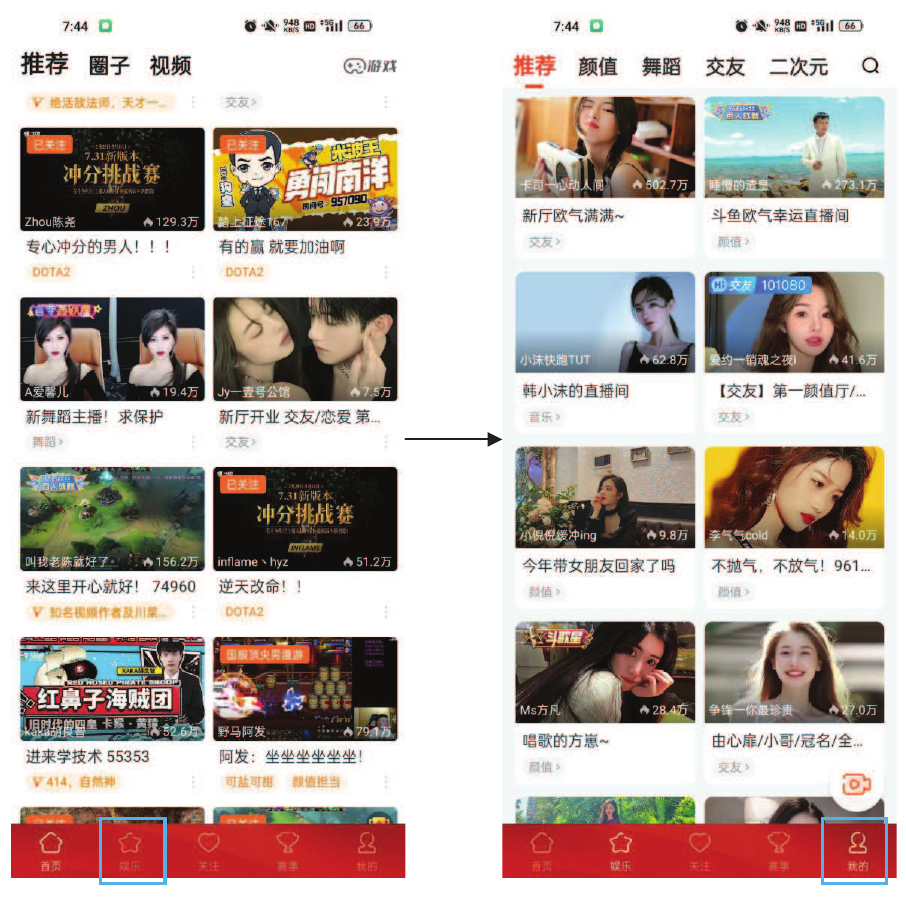}
  \caption{The UIs in Case 1 (best view in color).}
  \label{fig:Case1_UI}  
		\end{minipage}%
		\begin{minipage}[t]{0.45\linewidth}
			\centering
			
   \includegraphics[width=0.87\linewidth]{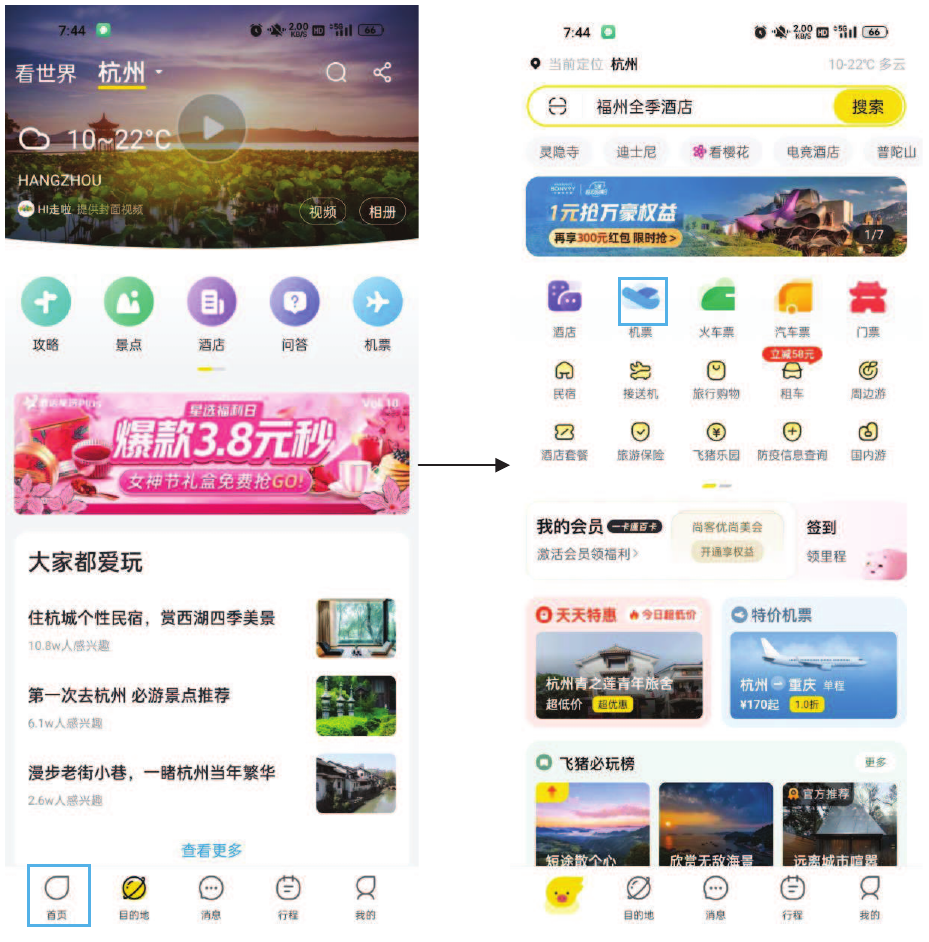}
  \caption{The UIs in Case 2 (best view in color).}
  \label{fig:Case2_UI}  
		\end{minipage}%

  \Description{}
\end{figure}
\begin{figure}[h]
\centering    
    \subfigure[SHA-SCP] 
    {
     \begin{minipage}{0.4\linewidth}
     \centering          
     \includegraphics[scale=0.4]{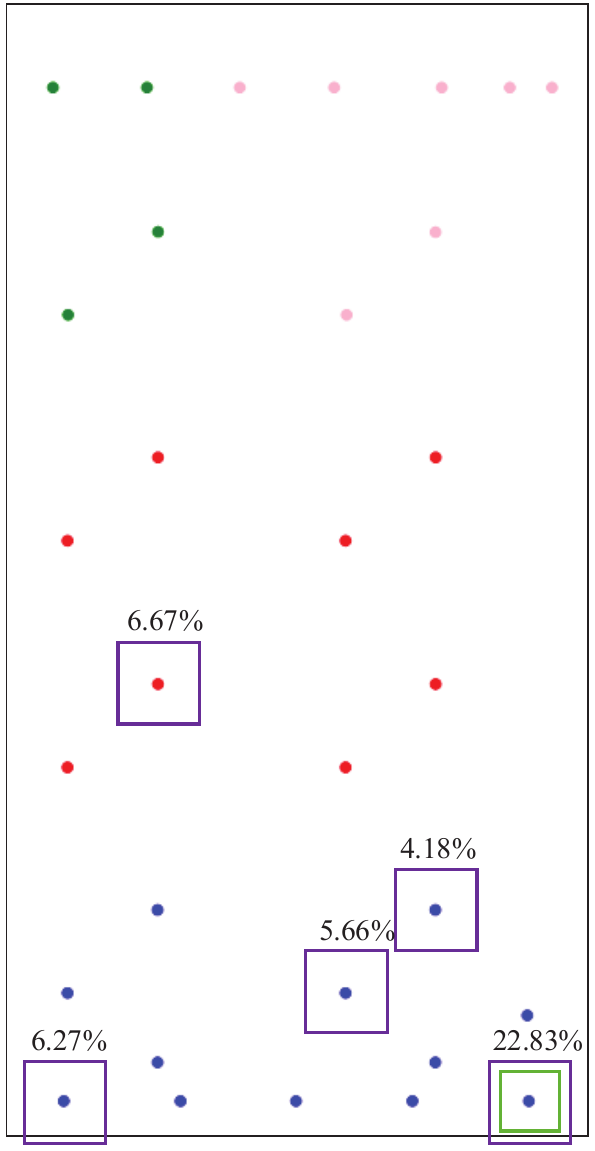}
     \end{minipage}
    }
    \subfigure[Google Model] 
    {
     \begin{minipage}{0.4\linewidth}
     \centering      
     \includegraphics[scale=0.4]{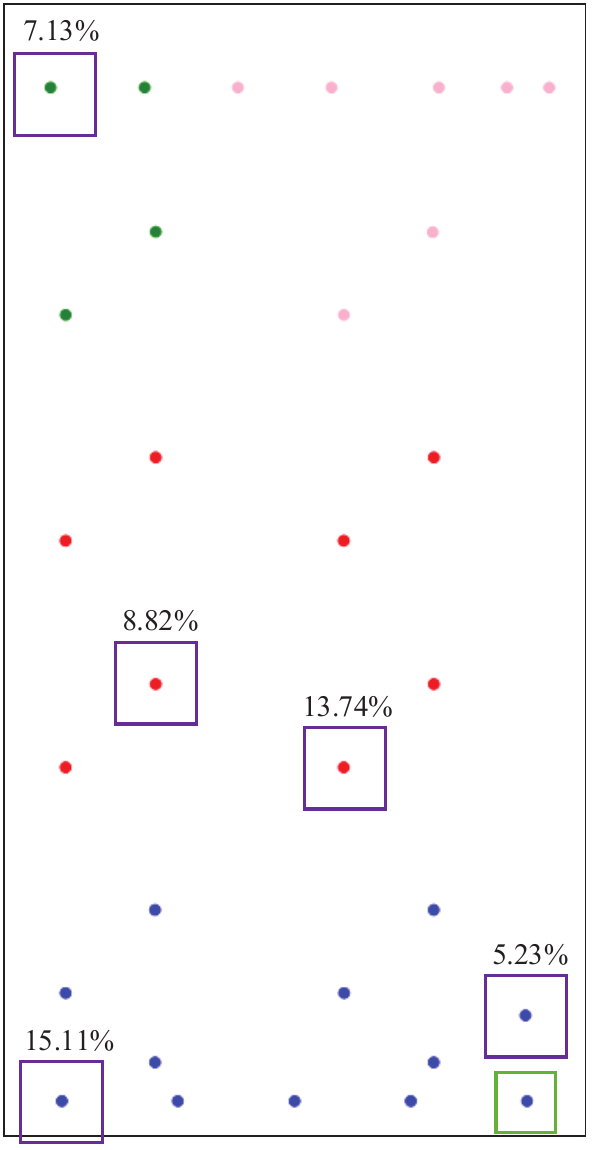}
     \end{minipage}
    }
\caption{The visualization of prediction results for the second click in Case 1 (best view in color).} %
\label{fig:Case1}  
\end{figure}
To intuitively reveal the superiority of SHA-SCP, we collect some fine-grained smartphone usage behavior data with the screenshots of UIs to perform case studies. Fig. \ref{fig:Case1_UI} shows a case of the UIs of a click sequence of length 2, in which the elements in the blue box are the clicked elements. The first click is on the ‘Entertainment’ button of the navigation bar, and the second click is on the ‘Me’ button of the navigation bar. The Top-5 prediction results of SHA-SCP and Google Model for the second click are visualized in Fig. \ref{fig:Case1}. Scatters with the same color indicate that corresponding elements belong to the same element group in SHA-SCP. The element in the green box is the real clicked element, and the elements in the purple boxes are the Top-5 elements predicted to be clicked. The percentages above purple boxes are clicking probabilities of corresponding elements. We can see that SHA-SCP predicts correctly (the probability of the real clicked element is highest) and Google Model predicts incorrectly. In addition, the Top-5 elements predicted by SHA-SCP are more concentrated, and 4 of them (including the real clicked element) belong to the same element group. In contrast, the Top-5 elements predicted by Google Model predicts are more dispersed, and only 2 of them belong to the element group of the real clicked element.

\begin{figure}[t]
\centering    
    \subfigure[SHA-SCP] 
    {
     \begin{minipage}{0.4\linewidth}
     \centering          
     \includegraphics[scale=0.4]{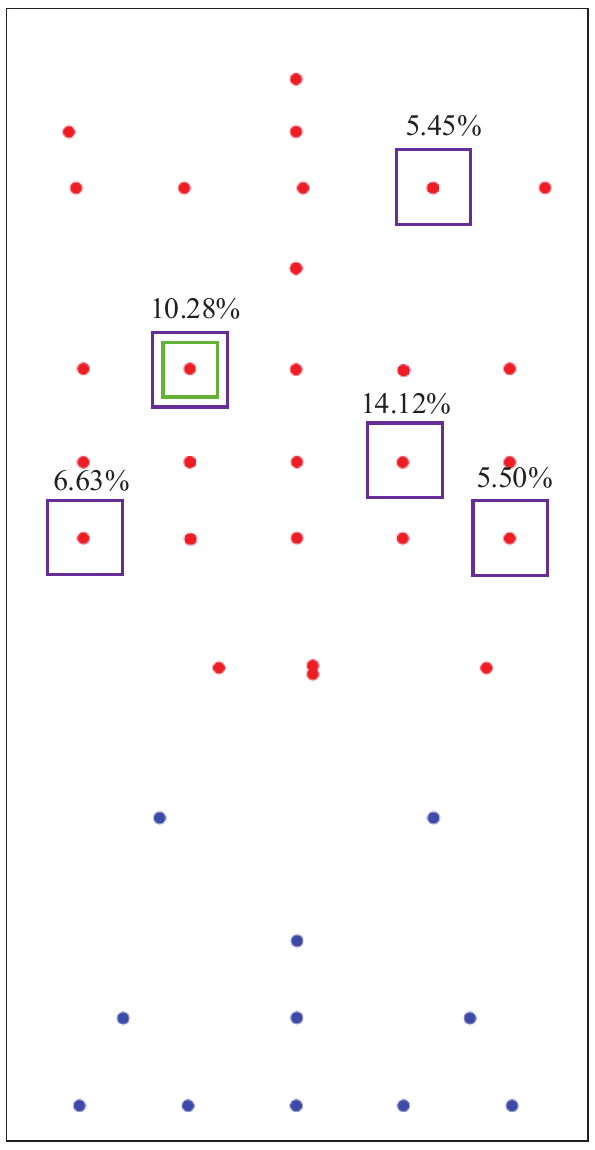}
     \end{minipage}
    }
    \subfigure[Google Model] 
    {
     \begin{minipage}{0.4\linewidth}
     \centering      
     \includegraphics[scale=0.4]{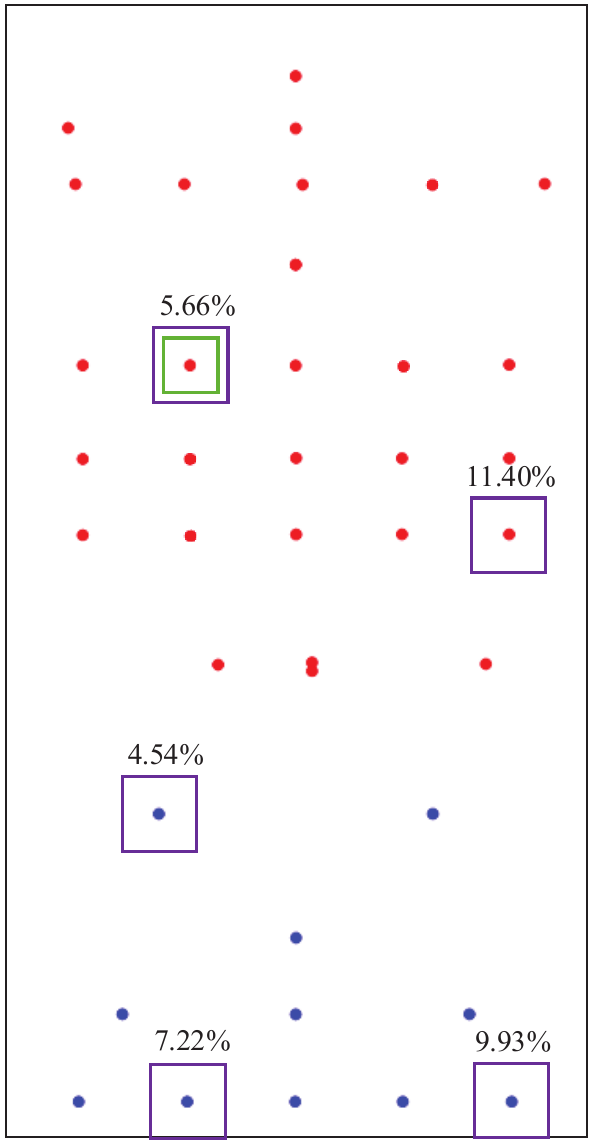}
     \end{minipage}
    }
\caption{The visualization of prediction results for the second click in Case 2 (best view in color).} %
\label{fig:Case2}  
\end{figure}

Fig. \ref{fig:Case2_UI} shows another case of the UIs of a click sequence of length 2. Similarly, the elements in the blue box are the clicked elements. The first click is on the ‘Home Page’ button of the navigation bar, and the second click is on the ‘Airline Tickets’ button in the middle of the UI. The Top-5 prediction results of SHA-SCP and Google Model for the second click are visualized in Fig. \ref{fig:Case2}. The meanings of colored scatters and boxes are the same as those in the previous case. It can be seen that SHA-SCP predicts correctly (the probability of the real clicked element is the second highest) and Google Model also predicts correctly (the real clicked element is ranked 4th). In addition, the Top-5 elements predicted by SHA-SCP are more concentrated, and the Top-5 elements predicted by Google Model are more dispersed.

Through the observation of two cases, we can find that the Top-5 elements predicted by SHA-SCP are more likely to be concentrated and belong to the same element group as the real clicked element, which further justify that by perceiving the UI at the element group level, SHA-SCP can effectively capture large-scale information in the UI for the user click behavior prediction task.

\begin{figure}[t]
    \subfigure[Element groups] 
    {
     \begin{minipage}{0.30\linewidth}
     \centering          
     \includegraphics[height=4cm]{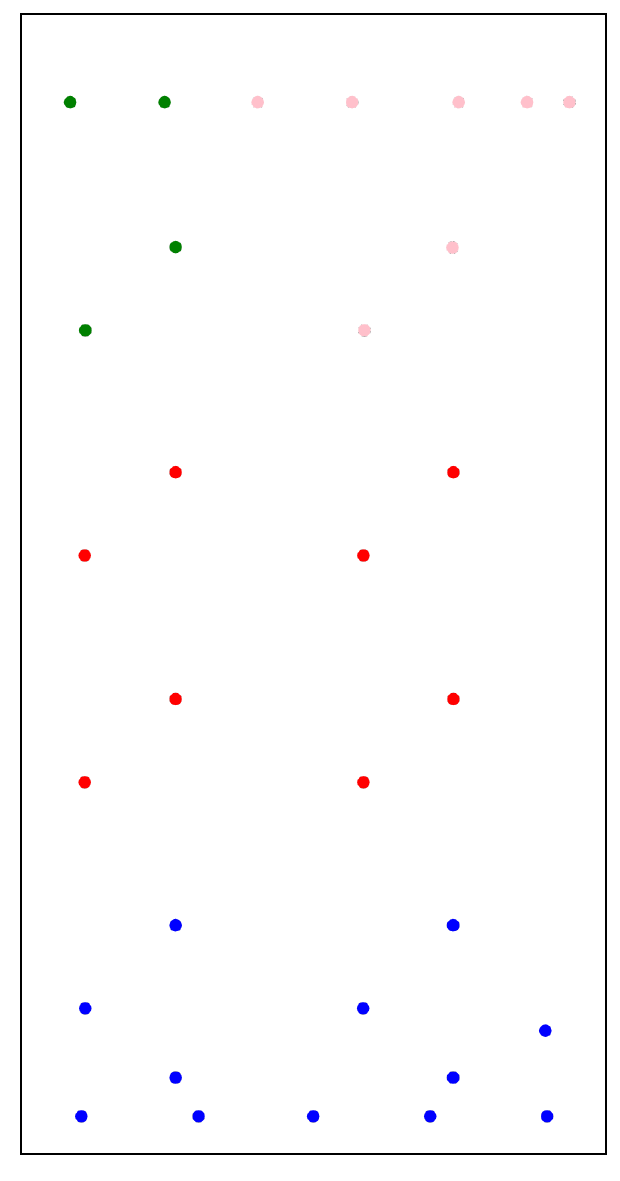}
     \end{minipage}
    }
    \subfigure[SHA-SCP] 
    {
     \begin{minipage}{0.30\linewidth}
     \includegraphics[height=4cm]{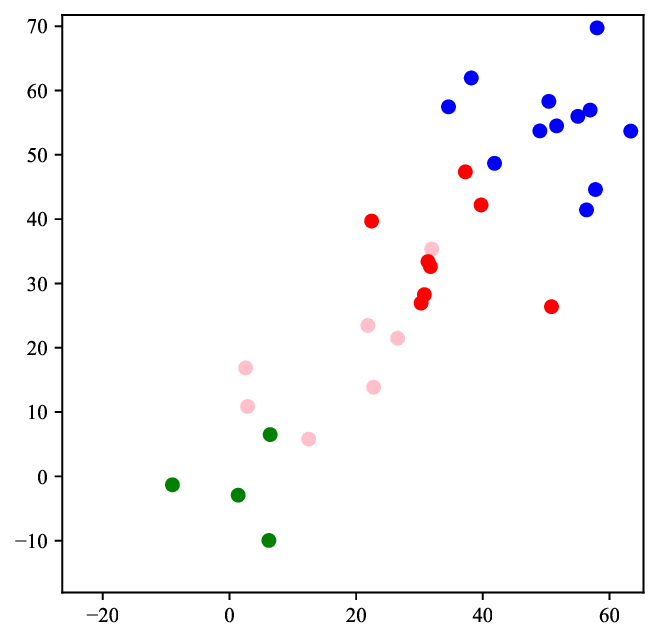}
     \end{minipage}
    }
    \subfigure[Google Model] 
    {
     \begin{minipage}{0.30\linewidth}
     \includegraphics[height=4cm]{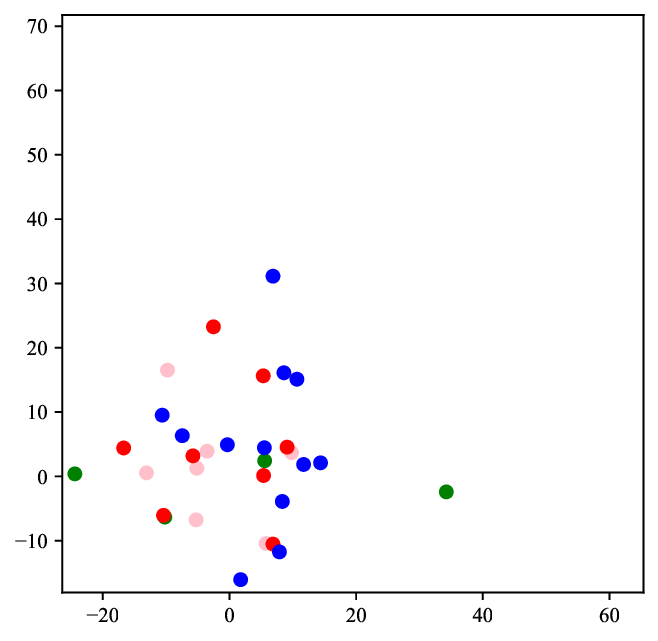}
     \end{minipage}
    }\hfill
    \caption{The visualization of the representations of elements in the UI (best view in color).} %
\Description{}
\label{fig:ele_present}  
\end{figure}

In addition, to further demonstrate the superior representations of elements provided by the element spatial hierarchy awareness and hierarchical element encoding modules, we visualize the representations of all elements in the UI of Case 1, comparing between SHA-SCP and Google Model. In Fig \ref{fig:ele_present} (b), it can be observed that elements, which are located closely in Fig \ref{fig:ele_present} (a), remain close after embedding. However, in Fig \ref{fig:ele_present} (c), due to the absence of the element spatial hierarchy awareness in Google Model, elements from the same group are seen to be farther apart, and elements from different groups, which should be separated, are seen to be mixed together. This further proves the effectiveness of exploiting the spatial hierarchy and clustering in SHA-SCP.

\section {LIMITATIONS AND FUTURE WORK}
In this section, limitations and future work related to our dataset and SHA-SCP are discussed. Firstly, current element groups are built using a distance-based method, which may not adapt to different user behaviors or UIs effectively, so we will explore other parameterized methods for building element groups. Secondly, our method is not efficient enough, so we will introduce the sparse attention mechanism with low complexity. Thirdly, considering the significance of contextual information in smartphone usage behavior prediction, we will collect a dataset that contains more contextual information, e.g., location. Fourthly, interaction with a smartphone consists of more than clicks on UI elements, but our method only focuses on predicting user click behaviors, so we will try to investigate if the model can also be used to predict other interactions. Lastly, our model is currently too large to run on smartphones, so we will investigate how to compress the model without sacrificing its prediction accuracy.

\section{CONCLUSIONS}

Smartphone usage behavior prediction is an important research area of mobile and ubiquitous computing. In this paper, we propose SHA-SCP, a UI element spatial hierarchy aware smartphone user click behavior prediction method, which extends the boundaries of this field by incorporating element spatial hierarchy awareness into the prediction of smartphone user click behaviors. By grouping elements based on their spatial positions and using attention mechanisms to perceive the UI at both the element and element group levels, SHA-SCP captures information at different scales, offering a more comprehensive UI perception for smartphone user click behavior prediction. In addition, we collect and analyze a fine-grained smartphone usage behavior dataset, on which comprehensive experiments, e.g., the comparison with baselines, the comparison with simplified models, the parameter sensitivity analysis, and the feature effectiveness analysis, are conducted and the results demonstrate the superiority of SHA-SCP. Furthermore, a case study is conducted, which further justifies the effectiveness of SHA-SCP.

\begin{acks}
This work is supported by the National Key Research and Development Program of China under Grant No.2018YFB0505000. 
\end{acks}

\bibliographystyle{ACM-Reference-Format}
\bibliography{sample-acmlarge}

\end{document}